\documentclass[sigconf,natbib=true,anonymous=false]{acmart}
\usepackage{color}
\usepackage{enumitem}

\usepackage{multirow}
\usepackage{bbding}
\usepackage{pifont}
\usepackage{subfigure}
\usepackage{caption}
\usepackage{tabularray}
\usepackage{booktabs}

\usepackage{subcaption}

\usepackage{amsmath}

\usepackage{amsthm,amssymb}

\usepackage{mathrsfs}

\AtBeginDocument{%
  \providecommand\BibTeX{{%
    \normalfont B\kern-0.5em{\scshape i\kern-0.25em b}\kern-0.8em\TeX}}}

\setcopyright{acmlicensed}
\copyrightyear{2018}
\acmYear{2018}
\acmDOI{XXXXXXX.XXXXXXX}

\acmConference[Conference acronym 'XX]{Make sure to enter the correct
  conference title from your rights confirmation emai}{June 03--05,
  2018}{Woodstock, NY}
\acmISBN{978-1-4503-XXXX-X/18/06}

\begin{document}
\newcommand{\minisection}[1]{\vspace{5pt}\noindent\textbf{#1.}}

\title{Look into the Future: Deep Contextualized Sequential
Recommendation}

\author{Lei Zheng}
\affiliation{%
  \institution{Shanghai Jiao Tong University}
  \city{Shanghai}
  \country{China}}
\email{zhenglei2016@sjtu.edu.cn}

\author{Ning Li}
\affiliation{%
  \institution{Shanghai Jiao Tong University}
    \city{Shanghai}
  \country{China}
}
\email{lining01@sjtu.edu.cn}

\author{Yanhua Huang}
\affiliation{%
  \institution{Xiaohongshu Inc.}
    \city{Shanghai}
  \country{China}
}
\email{yanhuahuang@xiaohongshu.com}

\author{Ruiwen Xu}
\affiliation{%
  \institution{Xiaohongshu Inc.}
    \city{Shanghai}
  \country{China}
}
\email{ruiwenxu@xiaohongshu.com}

\author{Weinan Zhang}
\authornote{Weinan Zhang is the corrpesponding author.}
\affiliation{%
  \institution{Shanghai Jiao Tong University}
    \city{Shanghai}
  \country{China}
}
\email{wnzhang@sjtu.edu.cn}

\author{Yong Yu}
\affiliation{%
  \institution{Shanghai Jiao Tong University}
    \city{Shanghai}
  \country{China}
}
\email{yyu@apex.sjtu.edu.cn}

\renewcommand{\shortauthors}{Trovato and Tobin, et al.}

\begin{abstract}

Sequential recommendation aims to estimate how a user's interests evolve over time via uncovering valuable patterns from user behavior history.
Many previous sequential models have solely relied on users' historical information to model the evolution of their interests, neglecting the crucial role that future information plays in accurately capturing these dynamics.
However, effectively incorporating future information in sequential modeling is non-trivial since it is impossible to make the current-step prediction for any target user by leveraging his future data.
In this paper, we propose a novel framework of sequential recommendation called Look into the Future (LIFT), which builds and leverages the contexts of sequential recommendation.
In LIFT, the context of a target user's interaction is represented based on i) his own past behaviors and ii) the past and future behaviors of the retrieved similar interactions from other users. As such, the learned context will be more informative and effective in predicting the target user's behaviors in sequential recommendation without temporal data leakage.
Furthermore, in order to exploit the intrinsic information embedded within the context itself, we introduce an innovative pretraining methodology incorporating behavior masking.
In our extensive experiments on five real-world datasets, LIFT achieves significant performance improvement on click-through rate prediction and rating prediction tasks in sequential recommendation over strong baselines, demonstrating that retrieving and leveraging relevant contexts from the global user pool greatly benefits sequential recommendation.
The experiment code is provided at https://anonymous.4open.science/r/LIFT-277C/Readme.md.
\end{abstract}



\settopmatter{printacmref=false}
\acmConference[KDD '25]{Proceedings of the 31th International ACM SIGKDD Conference on Knowledge Discovery \& Data Mining}{August 3--7, 2025}{Toronto, ON, Canada}
\renewcommand\footnotetextcopyrightpermission[1]{}



\maketitle

\section{Introduction}

Deep learning has been widely adopted for predicting user behaviors in online recommender systems \cite{zhang2021deep}. For the scenarios with sequences of user behaviors, sequential recommendation techniques, which extract valuable information from the past behavior sequence of the target user to predict his next behavior, have been well studied \cite{pi2019practice,huang2018improving}.

\begin{figure}[t]
    \centering
    \includegraphics[width=\linewidth]{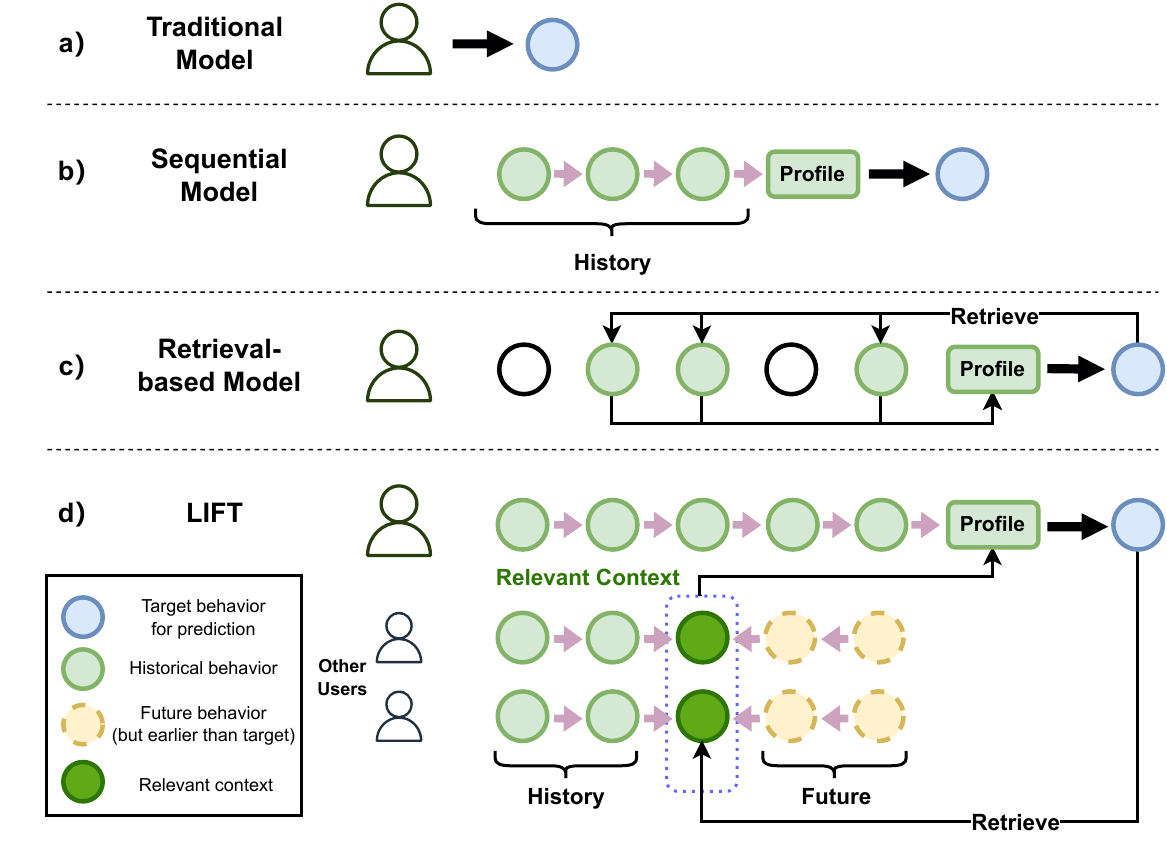}
    \caption{
    The comparison between LIFT and conventional models entails several key distinctions: a) Traditional models rely solely on instant user and item information when making predictions. b) Sequential models, conversely, typically incorporate the user’s historical interactions to capture their evolving interests over time. c) Retrieval-based models perform retrieval to fetch far-before but relevant historical behaviors to build the user profile for predictions. d) LIFT focuses on interaction context, encompassing both the historical and future sequence of interactions for each user-item interaction.
    %LIFT’s primary objective is to exploit this comprehensive context information to enhance prediction accuracy. However, note that in practical scenarios, accessing genuine future information is unattainable. To overcome this limitation, we leverage the future sequences of relevant observed interactions as a surrogate to approximate the future context for the current interaction.
}

\vspace{-10pt}

\label{liftfig:model-compare}
\end{figure}

As illustrated in Figure~\ref{liftfig:model-compare}, the major techniques for deep learning-based sequential recommendation are generally threefold. 
Firstly, deep architectures are designed to extend the capability of traditional collaborative filtering models for better mining of the feature interactions \cite{guo2017deepfm,zhang2016deep,wang2017deep}.
Secondly, different models like recurrent neural networks (RNN), memory networks, or transformers are adopted to learn an effective representation of the user behavior sequence, based on which a deep learning predictor is built to predict the behavior label for the current-step recommendation \cite{pi2019practice,chen2018sequential,sun2019bert4rec}.
Dealing with the long behavior sequence problem, thirdly, the retrieval methods are leveraged to fetch far early yet relevant behaviors, which are then aggregated according to the target prediction condition and fed into the final label predictor \cite{qi2020search, qin2020user}.

The primary aim of sequence modeling in recommender systems is to capture the evolving trends in user interests \cite{zhou2019deep}. Merely relying on historical data is insufficient to model these dynamic changes effectively \cite{yuan2020future}. Incorporating future information provides a hindsight perspective to help predict shifts in user preferences.
For instance, after purchasing a smartphone, a user is likely to show significant interest in related accessories. 
However, incorporating future data for predictive modeling without causing data leakage presents a significant challenge. 
Most existing models \cite{zhou2018deep, kang2018self} neglect the potential of future user behaviors, while a few attempts \cite{yuan2020future, sun2019bert4rec} implicitly use future information during training but fail to leverage it during the inference stage, leaving a significant potential for improvement. 

% This gap highlights the need for approaches that can effectively utilize future information to enhance the performance of recommender systems.

To address this issue, we employ retrieval-based methods to introduce context that contains future information into the model. Previously, retrieval techniques have been integrated into sequential recommendations to effectively access and leverage longer historical information \cite{pi2019practice,qin2020user}. Nonetheless, these retrieval methods typically focus on raw data from history, overlooking the importance of context. The interaction sequences of similar users with similar items tend to follow comparable patterns. By utilizing retrieval methods, we can fetch similar contexts in the log data from other users to approximate the future information of the target user. This approach eliminates the issue of temporal data leakage while enhancing the model's ability to predict user behaviors.

% However, the retrieval-enhanced models for sequential recommendation still retrieve in the ``raw'' data space, i.e., the individual behavior data, which actually loses the contextual information of the user making that particular behavior.
% Users' behaviors over the recommended items commonly exhibit interdependence.
% Typically, a user browses items with a specific intent, and within a certain time window, his attention is often drawn to items within a particular theme.
% For example, when a user is browsing ketchup, they may also browse many other kitchen condiments before and after.
% To this end, we posit the concept of a user's browsing sequence encompassing historical and future behaviors as the \emph{context} of that behavior.
% The incorporation of contextual information pertaining to the future behavior enhances the effectiveness of modeling user behaviors.
% Similarly, in the area of natural language processing (NLP),
% to model the word representation, the words before and after the target word are often used \cite{mikolov2013distributed, devlin2018bert}.
% Instead, in the field of recommender systems, the future behavior sequences are often overlooked \cite{zhou2018deep, zhou2019deep} because the future information is inaccessible when predicting the current behavior.
% However, we may obtain insight into future user behavior by identifying analogous contexts from the observed data of \emph{other users} and employing them to better model the context of the target user.

In this paper, we propose a novel framework of sequential recommendation called \underline{L}ook \underline{i}nto the \underline{F}u\underline{t}ure (LIFT), which focuses on retrieval in the user contextual information space while leveraging the future behaviors of the retrieved contexts, as illustrated on the bottom part of Figure~\ref{liftfig:model-compare}.
% LIFT works in scenarios where the behavior sequences of multiple users are accessible.
% In sequential recommendation, the representation of context interactions plays an important role in behavior prediction.
% All previous works in recommendation focus on modeling the historical part of the context.
% From a hindsight perspective, the entire context of the target interaction should include future interactions.
As mentioned above, it is infeasible to use real future information to predict the current behavior.
Thus, given the current-step context with the candidate item for behavior label prediction, LIFT performs retrieval to fetch the most similar interaction behavior from the whole user pool.
Then, extending from each of the retrieved behaviors, the historical and future behavior data are imported to enrich the context of that retrieved behavior.
Note that each of the included future behaviors is still earlier than the current timestep of the target user behavior, which avoids any data leakage issue.
Such future behaviors can be regarded as a kind of privileged information \cite{vapnik2015learning, xu2020privileged}, which is not accessible for current-step prediction but is accessible for historical behavior data.
To our knowledge, there is no previous work on sequential recommendation that performs effective retrieval of multiple users' behavior contexts that include both past and future behaviors.

% \vspace{-20pt}

Furthermore, it is crucial not only to utilize information derived from the target sample but also to exploit the inherent information within the context itself for the purpose of enhancing contextual representation.
Therefore, besides the supervised training with the label prediction loss, a representation learning loss is much important in our task.
To learn an effective representation of the context with both history and future behavior data, we further devise a pretraining method that performs masked behavior prediction during the pretraining stage.
Moreover, to reduce the noise introduced by the retrieved data, we designed an attention mechanism that assigns different weights to the retrieved data using dense embeddings of users and items.
We conduct extensive experiments on five real-world sequential recommendation datasets with click-through rate prediction and top-N ranking tasks, where LIFT demonstrates significant performance improvements over strong baselines.

Overall, the main contributions of this paper are threefold.
% \begin{itemize}
% \item We propose to incorporate the future information as part of the context information.
% We propose a novel framework that leverages retrieval techniques to use relevant context information to enhance the prediction performance of sequential recommendations while still avoiding data leakage.
% To our knowledge, this is the first work that makes the most use of the entire context information in the recommendation system based on retrieval techniques.

% \item In the pursuit of obtaining valuable representations of retrieved user behaviors, we adopt an approach that leverages intrinsic contextual information.
% Specifically, we devise a pretraining methodology that is tailored to the format of user sequence data and introduces a novel self-supervised loss function referred to as \emph{mask behavior loss}.

% \item We introduce a key-based attention aggregation technique aimed at consolidating contextual representations in accordance with the specific target sample and the keys contained within the retrieved contextual information.

% \end{itemize}

\begin{itemize}[leftmargin=10pt]
    \item We propose a novel LIFT framework that incorporates future information as part of the context. By leveraging retrieval techniques, LIFT utilizes relevant contextual information to enhance the prediction performance of sequential recommendations while avoiding temporal data leakage. To our knowledge, this is the first work to integrate future information comprehensively in both training and inference phases within a recommender system based on retrieval techniques.
    
    \item In the pursuit of obtaining valuable representations of retrieved user behaviors, we adopt an approach that leverages intrinsic contextual information.
    Specifically, we devise a pretraining methodology that is tailored to the format of user sequence data and introduces a novel self-supervised loss function referred to as \emph{mask behavior loss}.
    
    \item To avoid noise in the retrieved behavior sequences, we propose a key-based attention mechanism to aggregate the retrieved data.
    
\end{itemize}

\section{Related Work}
\minisection{Sequential Recommendation}
For sequential recommendation, user behavior modeling is the core technique that mines user preferences from historical interaction behaviors meticulously.
To better extract informative knowledge from user's behavior sequence, various network structures have been proposed, including Recurrent Neural Networks (RNNs), Convolutional Neural Networks (CNNs), Attention Networks, and Memory Networks.
GRU4Rec~\cite{hidasi2015session} designs Gated Recurrent Units (GRUs) to capture the preference-evolving relationship while Caser~\cite{tang2018personalized} leverages the horizontal and vertical convolution to model skip behaviors at both union-level and point-level.
Moreover, the attention mechanism is the most popular method for modeling item dependencies, and several influential works are proposed, including SASRec~\cite{kang2018self}, DIN~\cite{zhou2018deep}, DIEN~\cite{zhou2019deep}, and BERT4Rec~\cite{sun2019bert4rec}. 
% Among these, DIN and DIEN leverage the target-attention network to identify important historical items, while SASRec and BERT4Rec apply the self-attention network with Transformer architecture to excavate behavior dependencies.
% Besides, memory-based methods~\cite{wang2018neural,pi2019practice} are also proposed to store user behavior representations in a read-write manner. 
To better mine user preferences from their historical behaviors, two kinds of work are proposed to further extend the behavior modeling. The first one is multi-behavior modeling~\cite{zhou2018micro,yuan2022multi} that explicitly leverages different types of behavior (e.g., click and purchase behaviors) to measure item correlations within different behavior sequences, thus capturing users' diverse interests.
The second one is behavior modeling with side information~\cite{zhang2019feature} that involves various item attribute features (e.g., category) in addition to item IDs for better exploiting the rich knowledge.

\minisection{Retrieval-enhanced Recommendation}
To enhance the performance of recommender systems, the retrieval-enhanced recommendation is proposed, where the most relevant items are retrieved from an extremely long behavior sequence. Specifically, UBR4CTR~\cite{qin2020user} and SIM~\cite{qi2020search} are designed to retrieve beneficial behaviors from the user’s historical behavior, where UBR4CTR deploys the search engine method while SIM uses the hard search and soft search approaches.
To make the search procedure end-to-end, ETA~\cite{chen2021end} is proposed by leveraging the SimHash algorithm to map the user behavior into a low-dimensional space, hence achieving learnable retrieval. 
Moreover, recent works further extend the retrieval-enhanced recommendation from item-level retrieval to sample-level retrieval.
RIM~\cite{qin2021retrieval} is the first to deploy this method that leverages the search engine to retrieve several relevant samples from the search pool and performs neighbor aggregation.

\minisection{Pretraining for Recommendation}
Pretraining the deep learning models (or the data representation) with self-supervised learning methods has been widely studied in natural language processing and computer vision. Generally, there are two major categories of self-supervised training methods, namely contrastive learning \cite{gao2021simcse,chen2020simple} and mask recovery \cite{brown2020language,kenton2019bert,he2022masked}.
For recommender systems or tabular data prediction tasks, there are some recent attempts in this direction.
To list a few examples, BERT4Rec \cite{sun2019bert4rec} focuses on learning the representation of sequential behaviors via masking the final item of a subsequence of user behaviors.
SCARF \cite{bahri2022scarf} raises a contrastive learning loss via feature corruption over the tabular data.
MISS \cite{guo2022miss} proposes interest-level contrastive losses to
take the place of sample-level losses in order to
mine self-supervision signals from user behaviors of multiple interests.
S3\cite{zhou2020s3} proporsed to leverage inherent data correlations to generate self-supervision signals and improve data representations through pre-training methods.
However, these works borrow from the Cloze task in natural language processing by masking items to have the model predict item-related information, without considering the crucial role of behaviors in recommender systems.
Also, there are recent attempts to leverage the knowledge from the pretraining work on outsourced data, such as knowledge graphs \cite{wong2021improving} and language corpus \cite{cui2022m6}, to enhance the recommender systems.

\section{Formulation \& Preliminaries}\label{liftsec:pre}

In this section, we formulate the studied problem and introduce the preliminaries. 
A recommendation dataset $\mathcal{D}=\{d_z\}_{z=1}^N$ is represented as a set of interactions $d_z$ between user $u_z$ and item $i_z$. 
Each interaction $d_z$ can be formulated as the feature-label pair $(x_z, y_z)$, 
where $x_z = \{x_j^z\}_{j=1}^M$ contains the user and item features and a label $y \in \{0,1\}$ indicating whether the user will click the item or not, i.e., click-through rate (CTR).

A user's interaction sequence $s$ can be defined as a list of consecutive interactions for the same user sorted by time $s = [d_1, d_2, \ldots, d_T]$ where $d_1$ is the earliest interaction and $d_T$ is the latest one.
We define the last $L$ interactions before $d_z$ as the history sequence $h_z$ and the future $L$ interactions after $d_z$ as the future sequence $f_z$.
Then we define the full context $c_z$ as the combination of the history interaction sequence $h_z$ and future interaction sequence $f_z$:
\begin{equation}
    c_z = (h_z, f_z)~.
\end{equation}
To represent $d_z$, previous works on sequential recommendation only consider modeling the historical part of the context \cite{hidasi2015session,zhou2018deep}, while in this work, we use the full context to model $d_z$. 

The goal of a traditional recommendation task is to predict the label of the interaction $d_z$ based on the features $x_z$ from users, items, and the user $u_z$'s history behavior sequences. 
Such a prediction can be formulated as
\begin{equation}
    \hat{y}_z = F_\theta(x_z, h_z) ~,
\end{equation}
where $F_\theta$ is the learning model with the parameter $\theta$.

Here, we would use both the historical part information
$h_z$ and future part information $f_z$. However, the future information for $d_z$ is not visible in the inference stage. So we design a retrieval-based framework to retrieve the most relevant interactions' future sequence as the future sequence for the interaction $d_z$. Moreover, we still use the most similar interaction's history sequence to enhance the historical information.

Following the experiment setting from other retrieval-based sequential recommendation methods \cite{qin2021retrieval}, we split the dataset as $\mathcal{D}_{\text{train}}, \mathcal{D}_{\text{test}}, \mathcal{D}_{\text{retrieval}}$ for training, testing, and retrieval, respectively.
With the relevant contexts sequences, which include future sequences and history sequences, the label prediction is formulated as
\begin{equation}
    \hat{y}_z = F_\theta(x_z, h_z, C_{z})~,
\end{equation}
where $C_{z}$ represents the context sequences for $d_z$ and
\begin{equation}
C_z = R(x_z) ~,
\end{equation}
where $R$ is a retriever that could fetch relevant context sequences according to $x_z$.
We first retrieve relevant interactions $\{d_r\}$.
Then use the retrieved interactions to find their contexts to build $C_z$.
We will describe the procedure details in the next section.

Traditional methods only use the label from the target interaction $d_z$ to train $F_\theta$. 
However, if we only use the labels from the target interaction $d_z$, we will neglect most of the behavior information in the retrieved context sequences $C_z$.
Inspired by pretrained language models \cite{devlin-etal-2019-bert, brown2020language}, which have achieved tremendous success in the field of natural language processing (NLP) by learning universal representations in a self-supervised manner, we propose a self-supervised pretraining method that is supervised by the signal from the retrieved sequences itself. 
We denote the pretrained encoder as $E_{\omega}$ where the input of the encoder is any user interaction sequence $s$ and the output of $E_{\omega}$ is a $v$-dimension vector. 
The embedding of any user interaction sequence $s$ is written as
\begin{equation}
    \mathbf{s} = E_{\omega}(s)~,
\end{equation}
where $s \in \mathbb{R}^{1 \times v}$ and the retrieved context sequences set includes a history sequence set $H$ and a future sequence set $F$.
We retrieve $K$ context sequences, where $K$ can be tuned in this framework. Then the retrieved context embedding set is written as 
\begin{equation}
    \mathbf{C} = (\mathbf{H},\mathbf{F}) = (E_{\omega}(H), E_{\omega}(F))~,
\end{equation}
where $H \in \mathbb{R}^{K \times v}, F \in \mathbb{R}^{K \times v}, C \in \mathbb{R}^{K \times 2v}$.

As such, the parameters can be optimized by minimizing the self-supervised loss function as
\begin{equation}
    \omega^* = { \underset{ \omega }{{\arg\min}} \, \sum_{s \in \mathcal{S}_{\text{pretrain}}} \mathcal{L}_{\text{pretrain}}(E_\omega(s))}~,
\end{equation}
where $\mathcal{S}_{\text{pretrain}}$ is a sequence set sampled from the retrieved dataset $\mathcal{D}_{\text{retrieval}}$.
As the retriever $R$ is non-parametric, we optimize $F_\theta$ and $E_{\omega}$ independently. 
Specifically, after pretraining and fixing the parameter of $E_{\omega}$, we train $F_\theta$ via
\begin{equation}
    \theta^* = { \underset{ \theta }{{\arg\min}} \,  \sum_{(x_z, y_z) \in \mathcal{D}_{\text{train}}}   \mathcal{L}_{\text{prediction}}(F_\theta(x_z, \mathbf{h}_z, \mathbf{C}_z), y_z)}~,
\end{equation}
where $\mathbf{h}_z$ is the history sequence of $u_z$, encoded by $E_\omega$.

\begin{figure*}
    \centering
     \includegraphics[width=\linewidth]{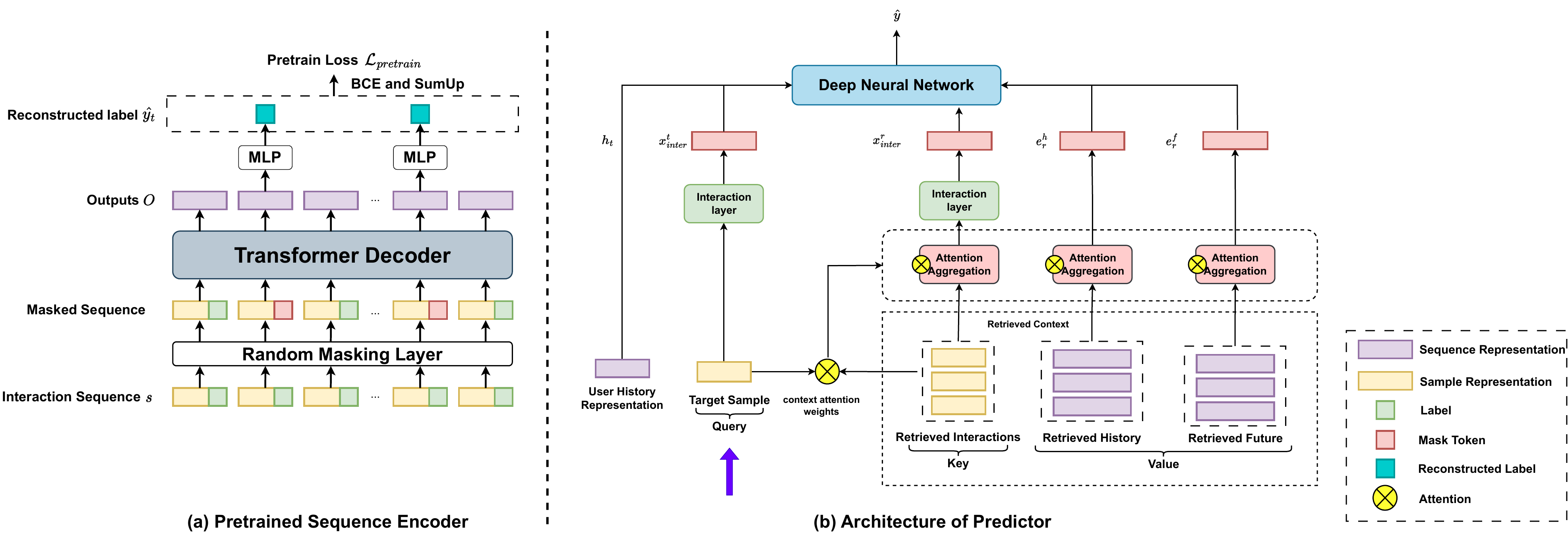}
     \vspace{-5pt}
    \caption{The architectural components of the encoder and predictor within the LIFT framework are as follows:
(a) Pretrained Sequence Encoder: The embedding layer is omitted in this component. LIFT employs a decoder-only Transformer architecture as the encoder, which undergoes pretraining via the mask behavior loss. During the pretraining stage, the primary focus is on leveraging contextual information inherent within the sequence data itself.
(b) Training of the Predictor: During the training stage, emphasis is placed solely on the training of the predictor. LIFT incorporates three distinct types of information to inform its predictions, namely the target sample $x_t$ itself, the user's historical interactions, and the retrieved context, encompassing historical interactions from similar instances as well as future interactions. In this phase, the label information utilized for training is exclusively derived from the target samples.
    }\label{liftfig:lift-encoder-predictor}
    \vspace{-0pt}
\end{figure*}

\section{The LIFT Framework}
In this section, we provide an overview of the whole LIFT framework and then describe the details of its the specific descriptions for its component encoder, retriever, and predictor, respectively.

\begin{figure}[h]
    \centering
    \includegraphics[width=1\linewidth]{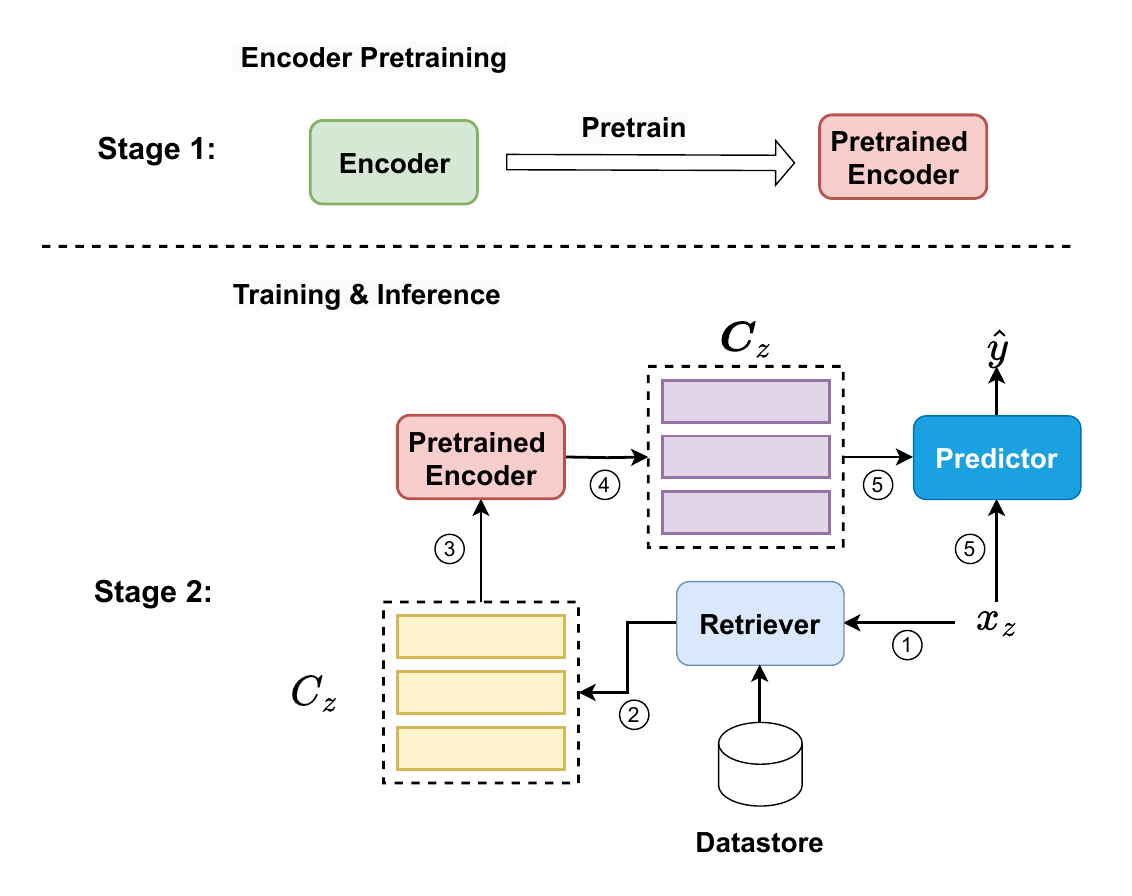}
    \vspace{-15pt}
    \caption{The overview workflow in LIFT. In the initial phase (Stage 1), we pretrain an encoder to convert sequences into embeddings. In the subsequent phase (Stage 2), 
    we illustrate the data flow in the figure. In Step 1, we send $x_z$  as a query to the retriever. In Step 2, the retriever outputs the retrieval result. In Step 3, we input the retrieval result into the encoder. In Step 4, we obtain the embeddings from the encoder. In Step 5, we send $x_z$ and the encoded embeddings to the predictor to obtain the final result.
    }

    % we establish a datastore utilizing the dataset $\mathcal{D}_{\text{retrieval}}$. Following this, we proceed to train the predictor denoted as "$Predictor$" using the dataset 
    % $\mathcal{D}_{\text{train}}$.
    \label{liftfig:components}
\end{figure}

\subsection{Overview}

There are three major components in LIFT, i.e., the encoder, retriever, and predictor. 
The encoder $E$ is responsible for encoding the user sequence to a $v$-dimensional vector. 
Given target interaction $d_t$, its feature part $x_t$ is regarded as a query, and the label part $y_t$ is to be predicted.
The retriever $R$ is used to retrieve relevant context sequences $C_t$. 
The predictor $F_\theta$ makes use of the feature vector $x_t$, the user $u_t$'s history sequence $h_t$, and the relevant context sequence set $C_t$ to predict the final label $\hat{y_t}$.

There are two steps in LIFT framework: \textit{pretraining} and \textit{predicting}. 
During pretraining, an encoder $E$ is trained on a  sampled $\mathcal{S}_{\text{pretrain}}$ from $\mathcal{D}_{\text{retrieval}}$ .
For $predicting$, the LIFT model first uses the retriever $R$ to find the top-$K$ relevant user context sequences $C_t$. 
And then, each sequence in $C_t$ will be encoded by $E_\omega$.
% For $predicting$, the LIFT model uses the pretrained encoder $E$ to encode all the target user behavior relevant sequences to help prediction. 
After aggregating the information of the target user and item $x_t$, the target user's history $h_t$, retrieved sequence $C_t$, the predictor $F_\theta$ will generate the output label prediction.

The main contribution of the LIFT framework is that it makes use of the full user context to help predict the target user behavior. Another distinctive feature of LIFT is that it leverages the pretraining method in a quite new paradigm and significantly improves the prediction performance in our experiment, which may open a new research direction of the sequential recommendation task.
Next, we will describe the three components of the LIFT framework in detail.

\subsection{Encoder}\label{liftsec:encoder} 
The encoder $E_\omega$ maps a user interaction sequence into a real-valued vector. It is pretrained separately in the framework.

\minisection{Model Architecture}
LIFT adopts a decoder-only Transformer as the encoder, which is based on the original implementation described in \citet{vaswani2017attention}. 
Transformer has been shown to perform strongly in language modeling \cite{brown2020language, devlin-etal-2019-bert}, machine translation \cite{vaswani2017attention}, recommender system \cite{sun2019bert4rec,zhang2024notellm}, etc. 
Different tasks may leverage different parts of the Transformer architecture. 
The difference between the Transformer encoder and decoder is that the self-attention sub-layer in the decoder is modified to make sure that the prediction in position $p$ could only depend on the positions earlier than $p$ \cite{vaswani2017attention}.
For the data of the recommender system exhibiting temporal characteristics, we use the decoder part of the Transformer.
We also compare the performance of different encoders in the experiment, which shows the superiority of the Transformer decoder.

\minisection{Input/Output} 
The input of the encoder $E_\omega$ is a user interaction sequence $s = \{ d_1, d_2, \ldots, d_T\}$. The output of $E_\omega$ is an embedding representing the sequence.

To feed the interaction $d_z$ into the encoder, $d_z$'s feature part $x_z$ is fed into an embedding layer. 
The embedding layer will map the one-hot encoded vector $x_z$ into a dense vector $\bf{x}_z$, where each feature $a_i$ in $x_z$ will be mapped into a $w$-dimensional  vector and then be concatenated into one $(M \times w)$-dimensional dense vector. 
The label part of the $d_z$ will be preserved. 
In the training stage, if the interaction $d_z$ is chosen to be masked, the label will be replaced by a special token \texttt{[MSK]}.
We will describe the training loss in the next subsection.
After processing the sequence, the output will be sent into the decoder-only Transformer decoder sequentially, and the output is a set of hidden states $O = \{o_1, o_2, ..., o_T\}$.
Because only the last step in the decoder can utilize the full information in the sequence, we choose the final state $o_T$ as the representation of the interaction sequence $\mathbf{s}$.

\minisection{Pretraining Data Preparation} 
The pretraining dataset $\mathcal{S}_{\text{pretrain}}$ is sampled from $\mathcal{D}_{\text{retrieval}}$.
For each user $u$ in $\mathcal{U}$, we use all the interactions in $\mathcal{D}_{retrieval}$ of $u$ to assemble an interaction sequence $s_u$. Then we get a sequence set $S$. For any sequence $s$ in $S$, we pick up a subsequence every $L$ interactions.
Using these subsequences, we get the pretraining dataset $\mathcal{S}_{\text{pretrain}}$.

\minisection{Mask  Behavior Loss} 
To leverage the inherent information contained within the context, we formulate a self-supervised loss function.
In contrast to certain existing sequential recommendation models \cite{zhou2018deep, zhou2019deep}, wherein only the target interaction label is considered as the prediction target for any given interaction sequence, all other interaction labels within the sequence are omitted.
It is rational to argue that incorporating more behavioral information from the context as the supervision signal would lead to a more comprehensive representation of the context. 
% Consequently, such an approach is expected to yield greater benefits for the prediction of target behavior.
And compared to previous pretraining methods \cite{sun2019bert4rec, zhou2020s3} that mask items or item parts, we propose masking user behavior to better incorporate user behavior information into the pretrained model, enhancing user-item interaction modeling.

% As shown in Figure \ref{liftfig:lift-encoder-predictor}, the left part describes the encoder architecture and loss in the pretraining procedure.
% In order to train a deep representation of the sequence, we mask part of the interaction's labels and then predict those masked behavior.
% Unlike the Cloze task \cite{taylor1953cloze} in natural language processing, it is difficult to predict the masked tokens in the user interaction sequences because there is no syntax in the recommendation. 
% An interaction $d_i$ in the input sequence $s$ can be represent as

As shown in Figure \ref{liftfig:lift-encoder-predictor}, the left part describes the encoder architecture and loss in the pretraining procedure. 
In order to train a deep representation of the sequence, we mask parts of the interaction's labels and then predict those masked behaviors. 
Unlike the Cloze task \cite{taylor1953cloze} in natural language processing, it is difficult to predict the masked tokens in the user interaction sequences because the data in recommendation systems lacks syntactic structure like that found in natural language. 
An interaction $d_i$ in the input sequence $s$ can be represented as:
\begin{equation}
d_i =
\left\{
\begin{matrix}
(x_i, \texttt{MSK}) & \text{if $d_i$ is a masked behavior}  \\
(x_i, y_i) & \text{otherwise} 
\end{matrix}
\right .
\end{equation}
where $i \in [1, T]$.
% In our experiments, we test different mask ratios, which shows that the best mask ratio should be different for different datasets. 

We adopt the binary cross entropy loss as the objective function. In the pre-training procedure, we feed the masked interactions' corresponding outputs as input into a multi-layer perceptron (MLP) to predict the masked labels. 
Thus the pretrain loss $\mathcal{L}_{\text{pretrain}}$ can be represented as $\mathcal{L}_{\text{pretrain}} = $
% \lei{Notions $d_i $is masked}\weinan{revised}
\begin{equation}
\begin{aligned}
        -\sum_{d_i \in \mathcal{M}} \left [  y_i \log\sigma(\text{MLP}(o_i))
    + (1 - y_i)\log(1 - \sigma(\text{MLP}(o_i))) \right]
\end{aligned}
\end{equation}
where $\mathcal{M}$ is the set of masked interactions and $\sigma(q)=1/(1+e^{-q})$ is the sigmoid function. We apply the $L_{\text{pretrain}}$ in every sequence in $S_{\text{pretrain}}$.In our experiments, we test different mask ratios, which shows that the best mask ratio should be different for different datasets. 
In contrast to the mask item loss\cite{sun2019bert4rec}, when dealing with sequences that are annotated with multiple behaviors, it becomes relatively more feasible for the mask behavior loss to propagate and encompass these labeled behaviors.

\subsection{Retriever}\label{liftsec:retriever}
The retriever $R$ is responsible for retrieving context sequences similar to the target interaction $d_t$. We use the feature part $x_t$ of $d_t$ as the query to utilize the search engine in finding similar user contexts in the retrieval pool. 
This section will cover the details of the datastore and search processes.

% We define the context $c_z$ for interaction $d_z$ as the history sequence $h_z$ that contains the $L$ interactions before $d_z$ and the future sequence $f_z$ that contains $L$ interactions after $d_z$ for the same user $u$. \lei{May simplified using definition in formula 1}
\minisection{Datastore} The datastore serves as a database in which the keys correspond to all the interactions within the retrieval dataset denoted as $\mathcal{D}_{\text{retrieval}},$ while the values correspond to their associated context sequences.
For the sake of efficiency, we employ a pretrained encoder to perform offline encoding of all these sequences into their respective embedding representations while the key is still the sparse format for retrieval.

The encoder $E_\omega$ encodes an interaction $d_z$'s context $c_z$'s history part $h_z$ and future part $f_z$ into two fixed-length vector representations respectively. 
Thus for any sample $d_z$ in $\mathcal{D}_{\text{retrieval}}$, we define the key-value pair $(k_z, v_z)$, where the key $k_z$ is the raw features $x_z$ of $d_z$ and the value $v_z$ is the encoded context embedding that contains a history sequence embedding $\mathbf{h}_z$ and a future sequence embedding $\mathbf{f}_z$.
% As shown in Figure \ref{liftfig:data_store}, 
The datastore $(\mathcal{K}, \mathcal{V})$ is the set of all key-value pairs constructed from all the samples in the retrieval dataset $\mathcal{D}_{\text{retrieval}}$:
\begin{equation}
    (\mathcal{K}, \mathcal{V}) = \{(x_z, (E_\omega(h_z), E_\omega(f_z))) | d_z \in \mathcal{D}_{\text{retrieval}} \}~.
\end{equation}

\minisection{Query}
The retrieval process uses the target sample's raw features $x_t$ as the query to find the nearby keys in the datastore. 
We use the traditional BM25 \cite{robertson1995okapi} algorithm to retrieve relevant keys, and return the top $K$ ranked key-value pairs. 
In the BM25 algorithm, we treat any sample $x_z$ as the query and a key $x_d$ in the datastore as the document. As such, the ranking score can be calculated as
\begin{align}
    \text{RankScore}(x_z, x_d) &= \sum_{j=1}^M \text{IDF}(x_j^z)\cdot \mathbf{1} (x_j^z = x_j^d) ~,\\
    \text{IDF}(x_j^z) &= \log\frac{N-N(x_j^z) + 0.5}{N(x_j^z) + 0.5}~,
\end{align}
where $\mathbf{1}(\cdot)$ is the indicator function, $N$ is the number of data samples in $\mathcal{D}_{\text{retrieval}}$, and $N(x_j^z)$ is the number of data samples that have the categorical feature value $x_j^z$.

Let $\hat{\mathcal{K}}$ represent the retrieved keys and $\hat{\mathcal{V}}$ be the retrieved values from the retrieved set.
After the retrieval process, we obtain the top-$K$ relevance pairs in the datastore.

\subsection{Predictor}
After the retrieval process, we obtain the context of the target sample $x_t$. The contextual information includes the user $u_t$'s history and both the history and future information of the target sample's relevant interactions. 
As shown in the bottom part of stage 2 in Figure \ref{liftfig:components}, in the predictor, we use both the contextual information and the embedding of $x_t$ to produce the final label prediction $\hat{y_t}$.

\minisection{Key-based Attention Aggregation}
From the retriever, we get the context embedding set $\hat{\mathcal{V}}$ of relevant interactions.
Before feeding this to the final prediction layer, we use the key-based attention mechanism to aggregate the retrieved $\hat{\mathcal{K}}$ and $\hat{\mathcal{V}}$.
Let $x_i$ be the $i$-th sample in the key set $\hat{\mathcal{K}}$.
We use an embedding layer to convert $x_i$ into a $M \times w$-dimensional dense vector $\mathbf{x}_i$ and we use the same embedding layer to map $x_t$ into a $M \times w$-dimensional dense vector $\mathbf{x}_t$.
The key attention weight is defined as
\begin{equation}
    \alpha_i = \frac{e^{\mathbf{x}_i^T {W} \mathbf{x}_t}}{\sum_{j=1}^K e^{\mathbf{x}_j^T {W} \mathbf{x}_t}}~,
\end{equation}
where ${W}\in \mathbb{R}^{Mw \times Mw}$ is the attention layer parameter matrix.
The allocation of attention weights to value sets is determined based on their respective key attention weights.
Thus we could use $\bf{\alpha}$ to aggregate the retrieved key set $\hat{\mathcal{K}}$ and value set $\hat{\mathcal{V}}$.
For the retrieved samples in $\hat{\mathcal{K}}$, the aggregated vector $\mathbf{x}_r$ can be written as
$\mathbf{x}_r = \sum_{i=1}^K \alpha_i \cdot \mathbf{x}_i,~~ x_i \in \hat{\mathcal{K}}$.
For the history part of the retrieved context embeddings $\hat{\mathcal{V}}$, the aggregated embedding $\mathbf{e}_r^h$ can be written as $\mathbf{e}_r^h = \sum_{i=1}^K \alpha_i \cdot \mathbf{h}_i,~~ \mathbf{h}_i \in \hat{\mathcal{V}}$.
% \begin{equation}
%     \mathbf{e}_r^h = \sum_{i=1}^K \alpha_i \cdot \mathbf{h}_i,~~ \mathbf{h}_i \in \hat{\mathcal{V}}~.
% \end{equation}
Similarly, for the future part of the retrieved context embeddings $\hat{\mathcal{V}}$, the aggregated embedding $\mathbf{e}_r^f$ can be written as
$\mathbf{e}_r^f = \sum_{i=1}^K \alpha_i \cdot \mathbf{f}_i,~~ \mathbf{f}_i \in \hat{\mathcal{V}}$.
% \begin{equation}
%     \mathbf{e}_r^f = \sum_{i=1}^K \alpha_i \cdot \mathbf{f}_i,~~ \mathbf{f}_i \in \hat{\mathcal{V}}~.
% \end{equation}

\minisection{Prediction}
In the final prediction layer, we first use a factor interaction layer to perform the inter-sample interaction for the target sample and the retrieved key set. 
Since these two vectors are categorical data, one could use high-order interaction to further explore the useful patterns, but this is out of the scope of this paper. 
Finally, we feed the embedding of the target sample, retrieved keys, and retrieved values into an MLP to make the label prediction.

After the embedding layer, each sample $x_z$ in $\mathcal{D}$ can be represented as a concatenation of $M$ feature embeddings:
\begin{equation}
    \mathbf{x}_z = [{\mathbf{a}_1^z},\ldots,{\mathbf{a}_M^z}]~,
\end{equation}
where $\mathbf{a}_i^z$ represents the $i$-the feature's embedding of the sample $x_z$.
We use an interaction layer to compute the combination of these features as
\begin{equation}
    {\mathbf{x}_{\text{inter}}^z} = [\text{inter}(\mathbf{a}_1^z, \mathbf{a}_2^z),\text{inter}(\mathbf{a}_1^z, \mathbf{a}_3^z),\ldots,\text{inter}(\mathbf{a}_{M-1}^z, \mathbf{a}_M^z)]~,
\end{equation}
where the $\text{inter}(\cdot)$ represents the interaction function. 
We could use different interaction functions such as inner product \cite{guo2017deepfm}, kernel product \cite{qu2018product}, or micro-network \cite{qu2018product}.
We use the same architecture of the interaction layer for the target sample $x_t$ and aggregated vector $\mathbf{x}_r$.
After $\mathbf{x}_r$ going through the interaction layer, we get the  embedding  $\mathbf{x}_{\text{inter}}^r$.
Finally, we concatenate all the vectors from the framework as
\begin{equation}
     \text{hidden} = \text{concat}[\mathbf{x}_{\text{inter}}^t, \mathbf{x}_{\text{inter}}^r, \mathbf{e}_r^f, \mathbf{e}_r^h, \mathbf{h}_t]~,
\end{equation}
where $\mathbf{h}_t$ is the user $u_t$'s history representation.
As such, the label final predicted $\hat{y}_z$ is written as
\begin{equation}
    \hat{y}_z = \sigma(\text{MLP}(\text{hidden})) ~.
\end{equation}

Given the prediction is a binary classification for CTR prediction, we use the binary cross-entropy loss between the label $y_t$ and prediction $\hat{y}_t$ for training. 
% as
% \begin{equation}
%     \mathcal{L}_{\text{prediction}}(\hat{y}_t, y_t) =  - y_t \log \hat{y}_t - (1 - y_t) \log (1 - \hat{y}_t)~.
% \end{equation}

\subsection{Time Complexity \& Speedup}
In this section, we provide a brief time complexity analysis of LIFT framework and then discuss the speedup feasibility. The detailed discussions are deferred to Appendix~\ref{appendix:complexity}.

\minisection{Time complexity}
LIFT consists of two major phases.
The first is the pretraining phase targeted at context sequences. In this phase, the dataset $D_{\text{retrieval}}$ is segmented into a set of sequences, each of which is characterized by a length $L$. As the operations on each sequence with a time complexity of $O(1)$, the time complexity of this phase is $O(|D_{\text{retrieval}}| / L)$.

The second phase lies in the LIFT model training and inference. Besides the neural network computation, the retrieval process is incorporated into the model, accounting for a time complexity of $O(F) + O(F \cdot |D_{\text{retrieval}}| / U) = O(F \cdot |D_{\text{retrieval}}| / U)$, where $U$ denotes the total number of unique features encountered in $D_{\text{retrieval}}$, $F$ denotes the number of features in the query.

\minisection{Speedup}
To enhance the efficiency of the retrieval process, we may adopt a method of storing the retrieval outcomes, eliminating the need to conduct the retrieval procedure during online inference. This approach shifts the retrieval process to an offline setting. For instance, in recommender systems, this method involves storing the retrieval results for all (or frequent) users, along with their respective recalled candidate items. This method represents a balance between the complexities of space and time, optimizing resource utilization by trading off storage space for processing speed.

\section{Experiments} 

This section starts with five research questions (RQs), which we
use to guide the experiments and discussions.\footnote{We provide the experiment code with running instructions on Anonymous GitHub (https://anonymous.4open.science/r/LIFT-277C/Readme.md).}

\begin{itemize}[leftmargin=25pt]

    \item [\textbf{RQ1}] Does LIFT achieve the best performance?
    \item [\textbf{RQ2}] Does future data benefit the final result?
    \item [\textbf{RQ3}] How does the proposed pretraining method have a positive impact on the prediction performance?
    % \item [\textbf{RQ4}] How do the hyperparameters ($L$ and $K$) influence the performance of LIFT?
    \item [\textbf{RQ4}] Do different encoders and mask label ratios have different impacts on context representation learning?
    \item [\textbf{RQ5}] How is the time efficiency of LIFT and whether is it potential to be deployed online?
% \item [\textbf{RQ3}] Does the label-aware interaction contribute to the prediction performance?
% \item [\textbf{RQ5}] What is the influence of the retrieval set size? [merged with RQ3]
\end{itemize}

\subsection{Datasets}
We evaluate the performance of LIFT by conducting experiments for CTR prediction tasks on three large-scale real-world datasets, i.e., 
Taobao, Tmall, 
and Alipay\footnote{https://tianchi.aliyun.com/dataset/x, where `x' is `649', `42', and `53' for Taobao, Tmall, and Alipay, respectively.}.
For top-N ranking, we utilize two widely-used public recommendation datasets: MovieLen\footnote{https://grouplens.org/datasets/movielens/1m/} and LastFM\footnote{http://ocelma.net/MusicRecommendationDataset/lastfm-1K.html}.
We count the number of instances, fields of these datasets, which is shown in Table \ref{lifttable:dataset}.

% begin{table}[t]
% \centering
% \caption{Dataset statistics.}\label{lifttable:dataset}
% \vspace{-10pt}
% \resizebox{0.7\columnwidth}{!}{
% \begin{tabular}{cccc}
% \toprule
% Datasets & \# Instances & \# Fields & \# Users\\
% \hline
% Taobao & 100,150,807 & 4 & 987,994\\
% Tmall & 54,925,331 & 9 & 424,170\\
% Alipay & 35,179,371 & 6 & 498,308\\
% \toprule
% \end{tabular}}
% % \vspace{-5pt}
% \end{table}

\begin{table}[h]
\caption{Dataset statistics}\label{lifttable:dataset}
\vspace{-10pt}
\small
\begin{tabular}{cccc}
\toprule
Datasets & \# Instances & \# Fields & Task\\
\hline
Taobao & 100,150,807 & 4 & CTR Prediction\\
Tmall & 54,925,331 & 9 & CTR Prediction\\
Alipay & 35,179,371 & 6 & CTR Prediction\\
Movielens-1M & 1,000,209 & 7 & Top-N Ranking\\
LastFM & 18,993,371 & 5 & Top-N Ranking\\
\toprule
% AZ-Elec & & \\
\end{tabular}
\vspace{-15pt}
\end{table}

% \begin{figure}[t]
%     \centering
%     \includegraphics[width=0.8\linewidth]{figure/Dataset.pdf}
%     \vspace{-10pt}
%     \caption{Dataset split format.}
%     \vspace{-15pt}
%     \label{liftfig:datasetsplit}
% \end{figure}
% according to figure \ref{liftfig:datasetsplit}

For dataset preprocessing, we follow the common practice in RIM\cite{qin2021retrieval} and DERT \cite{zheng2023dense}. We split each dataset into three parts, i.e., retrieval set, train set, and test set, based on the global timestamps. We select the hyperparameters using cross validation over the training set. Specifically, the retrieval set consists of the earliest data instances, the test set comprises the latest data instances, and the remaining intermediate data instances are allocated to the train set. For pretraining, we use the retrieval set to pretrain the encoder. For the non-retrieval baseline models, we merge the retrieval set and the training set as the final training set. 

\subsection{Evaluation Metrics} 
We choose the widely used metrics area under the ROC curve (AUC) and negative log-likelihood (LogLoss) to evaluate the performance for CTR prediction.
For top-N ranking, we use hit ratio (HR@N), normalized discounted cumulative gains (NDCG@N), and mean reciprocal rank (MRR). Significance test on each metric between the first and second performed methods is conducted, with * marked for positive test results.
We also conduct a significance test on the metrics between LIFT and the best performed baseline, marked with ``*'' for positive results. 

\begin{table*}[t]
\centering
\caption{Performance comparison of CTR prediction task baselines. GBDT and DeepFM are the traditional methods. Others are sequential modeling methods. For a fair comparison, traditional models are trained on both the retrieval set and the training set.
The best results are in bold fonts while the second best results are underlined.
``Rel. Impr.'' of each row means the relative AUC improvement of LIFT against the baseline. Improvements are statistically significant with $p < 0.01$.}
\vspace{-10pt}
\resizebox{0.8\textwidth}{!}{
\begin{tabular}{cccccccccc}
\toprule
\multirow{2}{*}{Models}  & \multicolumn{3}{c}{Taobao} & \multicolumn{3}{c}{Tmall} & \multicolumn{3}{c}{Alipay}   \\ \cline{2-10} 
    & LogLoss & AUC & Rel. Impr. & LogLoss & AUC & Rel. Impr.& LogLoss & AUC & Rel. Impr. \\
\hline
% LR &  &  &  &  &  &  &  &  &  \\
% PNN &  &  &  &  &  &  &  &  &  \\
GBDT & 0.6797 & 0.6134 & 44.39\% & 0.5103 & 0.8319 & 11.13\% & 0.9062 & 0.6747 & 30.49\% \\
DeepFM & 0.6497 & 0.6710 & 32.00\% & 0.4695 & 0.8581 & 7.74\% & 0.6271 & 0.6971 & 26.29\% \\
FATE & 0.6497 & 0.6762 & 30.98\% & 0.4737 & 0.8553 & 8.09\% & 0.6199 & 0.7356 & 19.68\% \\
BERT4Rec & 0.6356 & 0.6852 & 29.26\% & 0.4017 & 0.8981 & 2.94\% & 0.6024 & 0.7321 & 20.26\% \\
DIN & 0.6086 & 0.7433 & 19.16\% & 0.4292 & 0.8796 & 5.10\% & 0.6044 & 0.7647 & 15.13\% \\
DIEN & 0.6084 & 0.7506 & 18.00\% & 0.4445 & 0.8838 & 4.61\% & 0.6454 & 0.7502 & 17.36\% \\
SIM & 0.5795 & 0.7825 & 13.19\% & 0.4520 & 0.8857 & 4.38\% & 0.6089 & 0.7600 & 15.84\% \\
UBR & 0.5432 & 0.8169 & 8.42\% & 0.4368 & 0.8975 & 3.01\% & 0.5747 & 0.7952 & 10.71\% \\
RIM & {0.4644} & {0.8563} & 3.43\% & {0.3804} & {0.9138} & 1.17\% & {0.5615} & {0.8006} & 9.97\% \\
DERT & {0.4486} & {0.8647} & 2.42\% & {0.3585} & {0.9200} & 0.4\% & {0.5319} & {0.8087} & 8.86\% \\

% \hline
% LR + CL &  &  &  &  &  &  &  &  &  \\
% PNN + CL &  &  &  &  &  &  &  &  &  \\
% RIM + CL &  &  &  &  &  &  &  &  &  \\
\hline
% DERT(LR) &  &  &  &  &  &  &  &  &  \\
% DERT(LR + CL) &  &  &  &  &  &  &  &  &  \\
% DERT(PNN) &  &  &  &  &  &  &  &  &  \\
% DERT(PNN + CL) &  &  &  &  &  &  &  &  &  \\
LIFT w/o pretrain & \underline{0.4369*} & \underline{0.8727*} & 1.49\% & \underline{0.3509*} & \underline{0.9236*} & 0.10\% & \underline{0.4707*} & \underline{0.8572*} & 2.71\% \\

LIFT & \textbf{0.4129*} & \textbf{0.8857*} & - & \textbf{0.3489*} & \textbf{0.9245*} & - & \textbf{0.4361*} & \textbf{0.8804*} & - \\
\toprule
\end{tabular}}
\vspace{-8pt}
\label{lifttab:ctr}
\end{table*}

\begin{table*}[ht]
    \centering
    \caption{Performance comparison of top-N ranking tasks on the ML-1M and LastFM datasets. The best results are highlighted in bold, while the second-best results are underlined. Significant improvements are indicated by $p < 0.01$.}
    \label{lifttab:topn}
    \vspace{-10pt}
    \resizebox{0.7\textwidth}{!}{
    \begin{tabular}{llccccccccc}
        \toprule
        Datasets & Metric & FPMC & TransRec & NARM & GRU4Rec & SASRec & RIM & DERT & LIFT \\
        \midrule
        \multirow{6}{*}{ML-1M} 
        & NDCG@5 & 0.0788 & 0.0808 & 0.0866 & 0.0872 & 0.0981 & 0.1577 & \underline{0.1634} & \textbf{0.1806}$^*$ \\
        & NDCG@10 & 0.1184 & 0.1217 & 0.1254 & 0.1265 & 0.1341 & 0.2059 & \underline{0.2117} & \textbf{0.2293}$^*$\\
        & MRR & 0.1041 & 0.1078 & 0.1113 & 0.1135 & 0.1193 & 0.1704 & \underline{0.1774} & \textbf{0.1914}$^*$ \\
        & HR@1 & 0.0261 & 0.0275 & 0.0337 & 0.0369 & 0.0392 & 0.0645 &\underline{0.0747} & \textbf{0.0800}$^*$\\
        & HR@5 & 0.1334 & 0.1375 & 0.1418 & 0.1395 & 0.1588 & 0.2515 & \underline{0.2540} & \textbf{0.2808}$^*$\\
        & HR@10 & 0.2577 & 0.2659 & 0.2631 & 0.2624 & 0.2709 & 0.4014 & \underline{0.4035} & \textbf{0.4324}$^*$\\

        \midrule
        \multirow{6}{*}{LastFM} 
        & NDCG@5 & 0.0432 & 0.1148 & 0.0916 & 0.1229 & 0.1163 & 0.2165 & \underline{0.2620} & \textbf{0.2723}$^*$\\
        & NDCG@10 & 0.0685 & 0.1441 & 0.1185 & 0.1486 & 0.1409 & 0.2911 & \underline{0.3217} & \textbf{0.3444}$^*$\\
        & MRR & 0.0694 & 0.1303 & 0.1083 & 0.1362 & 0.1289 & 0.2210& \underline{0.2694} & \textbf{0.2727}$^*$\\
        & HR@1 & 0.0148 & 0.0563 & 0.0423 & 0.0658 & 0.0584 & 0.0915 & \textbf{0.1488}$^*$ & \underline{0.1485}\\
        & HR@5 & 0.0733 & 0.1725 & 0.1394 & 0.1785 & 0.1729 & 0.3468 & \underline{0.3742} & \textbf{0.4010}$^*$\\
        & HR@10 & 0.1531 & 0.2628 & 0.2227 & 0.2581 & 0.2499 & \underline{0.5780} & 0.5597 & \textbf{0.6310}$^*$\\
        
        \bottomrule
    \end{tabular}}
\end{table*}

\begin{table*}[htb]
\centering
\caption{
The influence of different encoders of LIFT on prediction performance. ``(p)'' means the encoder is pretrained.
%The influence of the retrieval mechanism for building the batch in contrastive learning of the encoder.
}
\label{lifttab:encoder}
\vspace{-10pt}
\resizebox{1.0\textwidth}{!}{
\begin{tabular}{ccccccccccccc}
\toprule
\multirow{2}{*}{Encoder} & \multicolumn{2}{c}{RNN} &\multicolumn{2}{c}{RNN (pretrain)} &  \multicolumn{2}{c}{Transformer Encoder} & \multicolumn{2}{c}{Transformer Encoder (p)} &  \multicolumn{2}{c}{Transformer Decoder} & \multicolumn{2}{c}{Transformer Decoder (p)}  \\ \cline{2-13} 
& LogLoss & AUC & LogLoss & AUC & LogLoss & AUC & LogLoss & AUC & LogLoss & AUC & LogLoss & AUC \\
\hline
Taobao & 0.4434 & 0.8667 & 0.4360 & 0.8715 & 0.4432 & 0.8684 & 0.4454 & 0.8679 & 0.4369 & 0.8727 & \bf{0.4129} & \bf{0.8857} \\
Tmall & 0.3730 & 0.9132 & 0.3637 & 0.9175 & 0.3539 & 0.9223 & 0.3528 & 0.9227 & 0.3509 & 0.9236 & \bf{0.3489} & \bf{0.9245} \\
Alipay & 0.4937 & 0.8391 & 0.4856 & 0.8468 & 0.4807 & 0.8504 & 0.4707 & 0.8572 & 0.4730 & 0.8556 & \bf{0.4361} & \bf{0.8804} \\
\bottomrule
\end{tabular}
}
\vspace{-5pt}
\end{table*}

\begin{table}
\centering
\caption{The influence of the usage of future data in LIFT.}
\label{lifttab:future}
\vspace{-10pt}
\resizebox{1.05\columnwidth}{!}{
\begin{tabular}{ccccccccc}
\toprule
\multirow{2}{*}{Dataset} & \multicolumn{2}{c}{No Context} &\multicolumn{2}{c}{Future Only} &\multicolumn{2}{c}{History Only} &  \multicolumn{2}{c}{Future \& History}   \\ \cline{2-9} 
& LogLoss & AUC& LogLoss & AUC & LogLoss & AUC & LogLoss & AUC \\
\hline
Taobao & .4644 & .8563 & .4417 & .8682 & .4405 & .8694 & \bf{.4129} & \bf{.8857} \\
Tmall & .3804 & .9138 & .3606 & .9191 & .3588 & .9199 & \bf{.3489} & \bf{.9245} \\
Alipay & .5615 & .8006 & .4781 & .8525 & .4780 & .8527 & \bf{.4361} & \bf{.8804} \\
\bottomrule
\end{tabular}
}
\vspace{-5pt}
\end{table}

\subsection{Compared Methods} 
In CTR prediction, we compare LIFT with nine solid baselines that can be categorized into three groups. The first group is traditional tabular models, which do not utilize the sequential or retrieval mechanism. This group includes GBDT \cite{chen1996data}, a widely-used gradient-boosted trees model, and DeepFM \cite{guo2017deepfm}, a factorization-machine based deep neural network. 
The second group consists of end-to-end sequential deep models, such as DIN \cite{zhou2018deep} and DIEN \cite{zhou2019deep} that are attention-based recurrent neural networks for CTR prediction, and BERT4Rec \cite{sun2019bert4rec} that is a Transformers-based sequential recommendation model. 
The third group includes retrieval-based models, such as SIM \cite{qi2020search}, UBR \cite{qin2020user}, RIM \cite{qin2021retrieval} and DERT\cite{zheng2023dense}. Additionally, FATE \cite{21fate} is a tabular data representation learning model that can be viewed as a random retrieval method for prediction.
% Additionally, FATE \cite{21fate} is a tabular data representation learning model, which accepts a random mini-batch of data instances as the entire input and predicts all its elements simultaneously, enabling interactions between the samples in the mini-batch. This approach can be viewed as leveraging a random retrieval method for prediction.

For top-N ranking, we compare LIFT against seven baseline models. FPMC \cite{rendle2010factorization} and TransRec \cite{he2017neural} are factorization-based approaches. The other baselines—NARM \cite{li2017neural}, GRU4Rec \cite{hidasi2015session}, SASRec \cite{khandelwal2019generalization}, RIM \cite{qin2021retrieval} and DERT \cite{zheng2023dense}—are recently proposed neural network models.

\subsection{Overall Performance (RQ1)}

% The overall performance comparison result is provided in Table\ref{lifttab:ctr}, where we make the following observations: (i) LIFT consistently outperforms all nine baselines. Specifically, compared to the best baseline RIM, LIFT achieves a relative improvement in AUC by 3.43\%, 1.17\%, and 9.97\% on Taobao, Tmall, and Alipay, respectively. This clearly demonstrates the effectiveness of context information learned and retrieved in the LIFT framework. (ii) Without pretraining, LIFT still consistently outperforms the best baseline, indicating that the raw information of future and history still yields a significant improvement in prediction performance. (iii) The retrieval-based methods, LIFT and RIM, are superior to the traditional methods and sequential CTR models, which suggests that the retrieval methods are able to make better use of the context information. (iv) In Table\ref{lifttab:topn}, it can be observed that LIFT achieves significant improvements over these baselines across nearly all metrics on both datasets. This demonstrates that incorporating contextual information enables the sequential model to perform well in top-N ranking tasks.

The overall performance comparison result is provided in Table~\ref{lifttab:ctr}, where
we have the following observations: 
(i) LIFT consistently outperforms all nine baselines. 
Specifically, compared to the best baseline RIM, LIFT
achieves a relatively improved AUC by 
3.43\%, 1.17\% and 9.97\% on 
Taobao, Tmall, and Alipay, respectively. This clearly demonstrates
the effectiveness of context information learned and retrieved in LIFT framework.
(ii) Without pretraining, LIFT still consistently outperforms the best baseline, 
which indicates that the raw information of future and history still yields a significant improvement on the prediction performance.
(iii) The the retrieval-based
methods LIFT and RIM are superior to the traditional methods
and sequential CTR models, which means that the retrieval
methods are able to make better use of the context information. 
(iv) in the table~\ref{lifttab:topn}, it can be observed that LIFT achieves significant improvements over these baselines across nearly all metrics on both datasets. 
This demonstrates that incorporating contextual information enables the sequential model to perform well in top-N ranking tasks.

\subsection{Further Analysis} 
We further study the effectiveness of important modules in LIFT, i.e., the usage of future context, the pertaining, the encoder architecture, and the hyperparameters.

\minisection{Future Context (RQ2)} 
Table~\ref{lifttab:future} compares the different parts of the context information's impact on the final prediction performance.
We can observe that both the history information and future information significantly impact on the final prediction. 
If we only use part of the context information, such as future only or history only, we get a worse result than using both of them, which indicates that the future part of the context provides different information compared with the history part. This observation is entirely neglected by previous works.

\minisection{Pretraining (RQ3)} From the last two rows of Table \ref{lifttab:ctr}, we find that over the three datasets, the model with pretraining performs better than those without, which clearly demonstrates the efficiency of pretraining.
The enhanced performance of the pretrained encoder implies that relying solely on the label provided by the target sample is insufficient for effectively capturing contextual information. 
To improve the modeling of context, it is imperative to harness the inherent information embedded within the context itself.

% , especially in the dataset Taobao and Alipay, which probably means that dataset with better sequential pattern will get benefit more in the pretraining procedure.

% \subsection{Hyperparameter Study} % (Lei Zheng)

% \minisection{Hyperparameter in the Retriever  (RQ4)} 
% In the LIFT framework, $L$ means the context sequence length and $K$ means the retrieved samples count. From Figure \ref{liftfig:taobao_hyper} and Figure \ref{liftfig:alipay_hyper}, we can find that with the increase of $K$, the final prediction result first goes up and then down. The curve indicates that along with with the $K$, the information first increases, and then more noises are introduced to become the dominant part. The AUC curves of $L$ show the similar trends of $K$, i.e., as $L$ increases, the AUC first gets better and then drops down.
% We think it is the same reason as for $K$.
% Moreover, because of the limitation of the GPU resources, we only conduct this hyperparameter study on Taobao and Alipay. and we could only increase the $L$ to 70 in the history length of Taobao, where the downtrend just started.

\begin{figure}[t]
    \centering
    \includegraphics[width=1\linewidth]{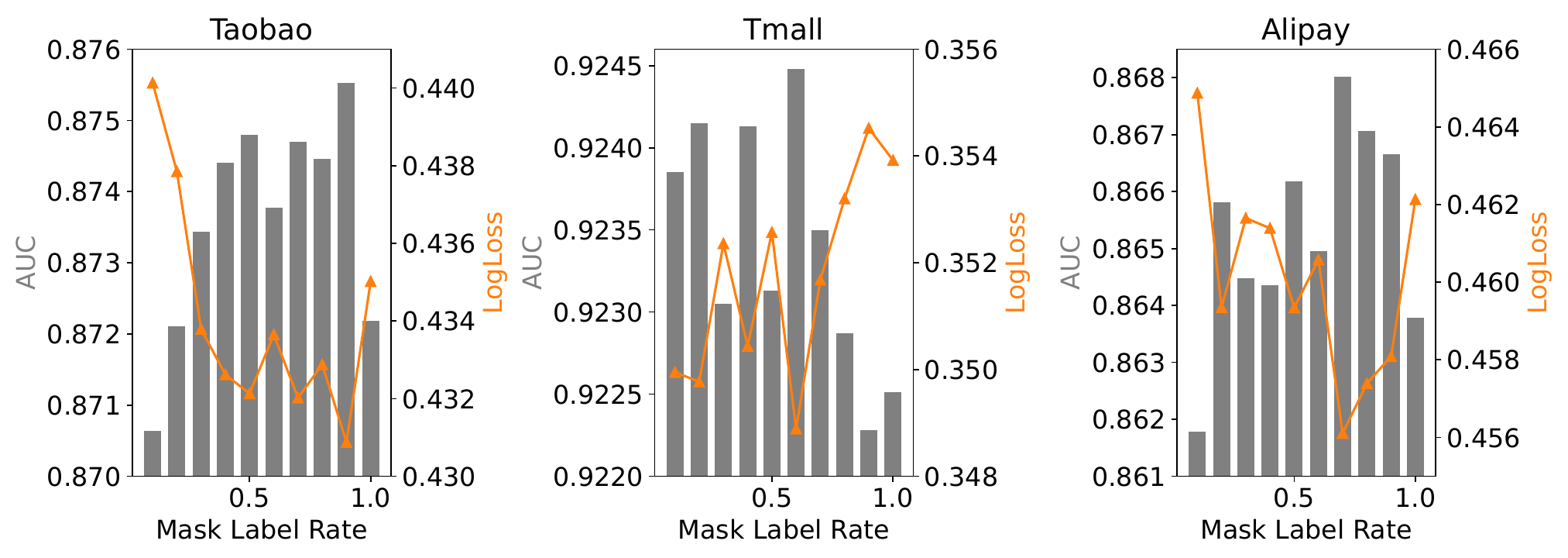}
% \vspace{-15pt}
\caption{Performance w.r.t. different pretraining mask rates.}
\vspace{-15pt}
\label{liftfig:mask_rate}
\end{figure}

\minisection{The Choice of the Encoder (RQ4)} 
The prediction performance of LIFT with different encoders is provided in Table~\ref{lifttab:encoder}, where we can observe that the Transformer decoder based encoder yields the best performance on both non-pretraining and pretraining settings. The reason could be that the Transformer decoder could be aware of the position of the sequence better. 
From Figure \ref{liftfig:mask_rate}, we observe the mask ratio's influence on the prediction performance. 
Usually, the traditional models that encode the user interaction sequence only use the last label as the supervised signal.
From the result, we can find that in all three datasets, such a $1$ label supervised strategy is not the best choice. 
The optimal mask label rate is different over datasets where the number is usually larger than 50\%.

\begin{figure}[htb]
    \centering
    \includegraphics[width=0.5\linewidth]{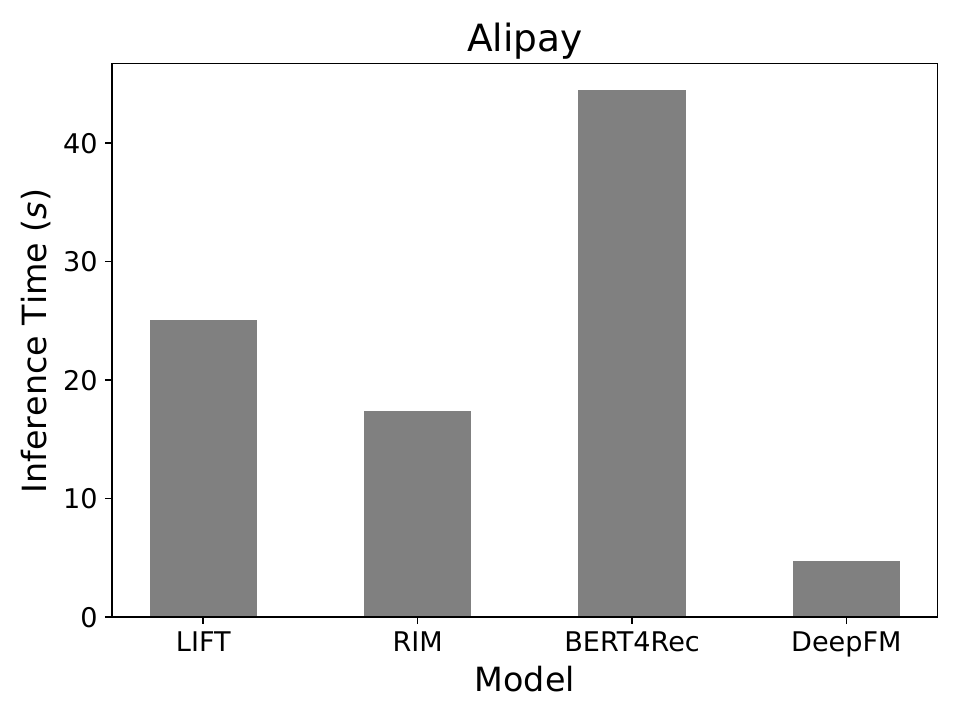}
\vspace{-15pt}
\caption{Inference time on Alipay.}%\vspace{-5pt}
\label{liftfig:inference_time}
\vspace{-15pt}
\end{figure}

\minisection{Time Efficiency (RQ5)} 
To evaluate the time efficiency of LIFT, we compare its real inference time against other three mainstream models in different types.
The parameters that performed best on Alipay were selected for each model to test the inference performance.
As shown in Figure~\ref{liftfig:inference_time}, LIFT does not introduce significant overload compared with the basic retrieval-based model RIM.
The optimal retrieval set size $K$ for LIFT's AUC on the Alipay dataset is 15. RIM with $K=10$ is approximately 30\% faster than LIFT. 
With offline pretraining, LIFT operates far more efficiently than the transformer-based model BERT4Rec.

\section{Conclusion} % (Lei Zheng)
In this work, we propose a retrieval-based framework called LIFT to better utilize the context information of the current user interaction. We are the first work to include future information as part of the context without temporal data leakage in the training and inference stages.
Moreover, we use a pretraining method to better mine the information in the context sequences and propose a novel mask behavior loss. 
The performance of the LIFT framework shows that both historical and future information yield significant improvements in the CTR prediction and top-N ranking performance.
Also, from the comparison of the pretraining method, we find that context pretraining is a promising solution to further improve the prediction performance of a variety of sequential recommendation models. 
In the future, we will investigate deeper on context representation learning with more sophisticated retrieval methods and
the speedup schemes to deploy LIFT to real-world recommenders.
\bibliographystyle{ACM-Reference-Format}
\bibliography{lift}

\clearpage

\appendix

\section{Notations}

The notations and their descriptions are summarized in Table \ref{lifttable:notation}.

\begin{table}[h]
\centering
\caption{Notations and corresponding descriptions.}\label{lifttable:notation}
% \vspace{-10pt}
\resizebox{1\linewidth}{!}{
\begin{tabular}{c|l}
\toprule
Notation & Description \\
\hline
$\mathcal{D}_{\text{train}}$, $\mathcal{D}_{\text{test}}$, $\mathcal{D}_{\text{retrieval}}$ & Training set, test set, retrieval set \\
% $\mathbb{D}_\cdot$ & The embeddings of the data instances in the corresponding set \\
$x_z$, $\mathbf{x}_z$ & The raw feature and the embedding of $z$-th sample\\
$x_t$, $\mathbf{x}_t$ & The raw feature and the embedding of target sample\\
$E_\omega$ & The encoder and its parameters \\
$R$ & The retriever \\
$F_\theta$ & The predictor and its parameters \\
$d$ & An interaction \\
$v$ & The encoder output vector dimension \\
$w$ & The feature embedding vector dimension \\
% $S$ & The set of retrieved data instances \\
% $\mathbb{S}$ & The set of retrieved data instance embeddings \\
$K$ & The size of retrieved contexts \\ 
$L$ & The length of history and future sequence \\ 
$c$, $h$, $f$ & The context sequence, history sequence, future sequence \\
$C$, $H$, $F$ & The context sequence set, history sequence set, future sequence set\\
\toprule
\end{tabular}}
\end{table}

\section{Datastore of LIFT}
\begin{figure}[h]
    \centering
    \includegraphics[width=1\linewidth]{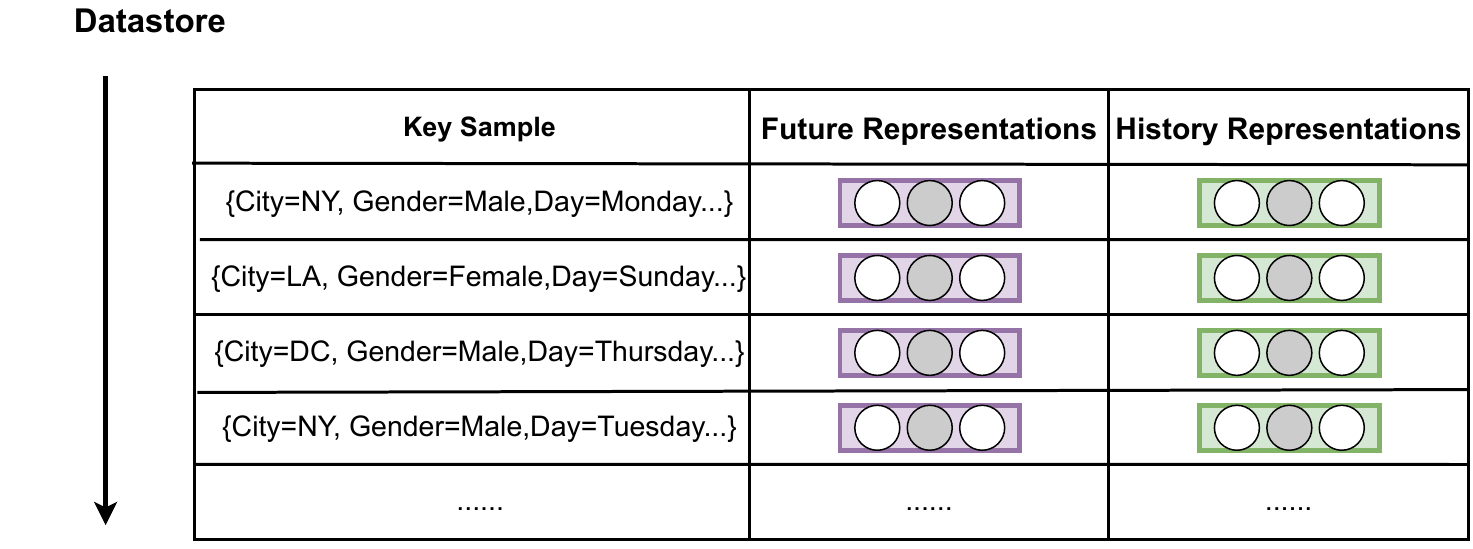}
    \caption{The datastore in LIFT. 
    }
    \label{liftfig:data_store}
    % \vspace{-10pt}
\end{figure}

As shown in Figure \ref{liftfig:data_store}, 
the datastore $(\mathcal{K}, \mathcal{V})$ is the set of all key-value pairs constructed from all the samples in the retrieval dataset $\mathcal{D}_{\text{retrieval}}$:
\begin{equation}
    (\mathcal{K}, \mathcal{V}) = \{(x_z, (E_\omega(h_z), E_\omega(f_z))) | d_z \in \mathcal{D}_{\text{retrieval}} \}~.
\end{equation}

The primary keys within this datastore are intricately linked to the interactions in the dataset $\mathcal{D}_{\text{retrieval}}$, which represents the original data format. 
In parallel, the datastore houses values that align with the contextual information associated with these interactions. 
It is noteworthy that this contextual information is derived through an encoding process that applies a pretrained encoder, denoted as $E$, to the context sequences.

\vspace{10pt}
\section{Time complexity}\label{appendix:complexity}
We aim to provide a comprehensive analysis of the time complexity within our proposed framework, which consists of two major components.

The first component addresses the pretraining phase targeted at context sequences. In this phase, the dataset $D_{\text{retrieval}}$ is segmented into a set of sequences, each of which is characterized by a length $L$. Both training and inference stages involve operations on each sequence with a time complexity of $O(1)$. Consequently, the overall time complexity during the pretraining phase is $O(|D_{\text{retrieval}}| / L)$.

The second component pertains to the complexity of the main framework's algorithm, which diverges from traditional neural network methodologies by incorporating a retrieval process during both training and inference phases. We commence by examining the complexity of this retrieval process. Here, $|D_{\text{retrieval}}|$ represents the quantity of samples, and $U$ denotes the total number of unique features encountered in $D_{\text{retrieval}}$.
Note that the mean length of the posting lists in the inverted index is $|D_{\text{retrieval}}| / U$. As explained in Section \ref{liftsec:retriever}, the retrieval operation, which encompasses retrieving all posting lists of features in $x_t$, necessitates a time complexity of $O(F)$, where $F$ symbolizes the number of features in the query. This phase is deemed a constant time operation. The average count of samples retrieved is $F \cdot |D_{\text{retrieval}}|/U$, and the complexity of the ranking operation scales linearly with the number of retrieved samples, indicated as $O(F \cdot |D_{\text{retrieval}}| / U)$. Therefore, the total time complexity of the retrieval process is $O(F) + O(F \cdot |D_{\text{retrieval}}| / U) = O(F \cdot |D_{\text{retrieval}}| / U)$.
Besides the retrieval process, the neural network computation time complexity of the predictor is  about the same with the other models such as RIM \cite{qin2021retrieval} or DIN \cite{zhou2018deep}. The detailed architecture acceleration would be an effective factor for the model inference effieiency.

\vspace{10pt}
\section{Hyperparameter in the Retriever}

\begin{figure}[h]
    \centering
    \includegraphics[width=1\linewidth]{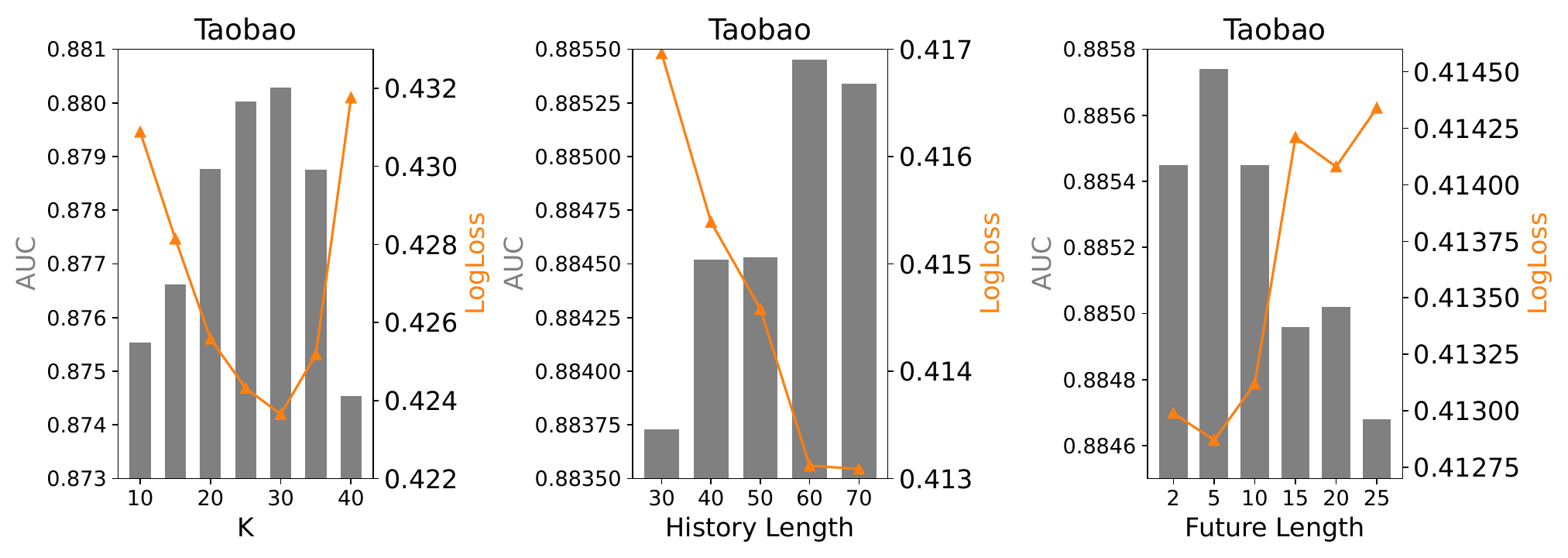}
% \vspace{-5pt}
\caption{Hyperparameters study of LIFT on Taobao.}%\vspace{-5pt}
\label{liftfig:taobao_hyper}
\end{figure}

\begin{figure}[htb]
    \centering
    \includegraphics[width=1\linewidth]{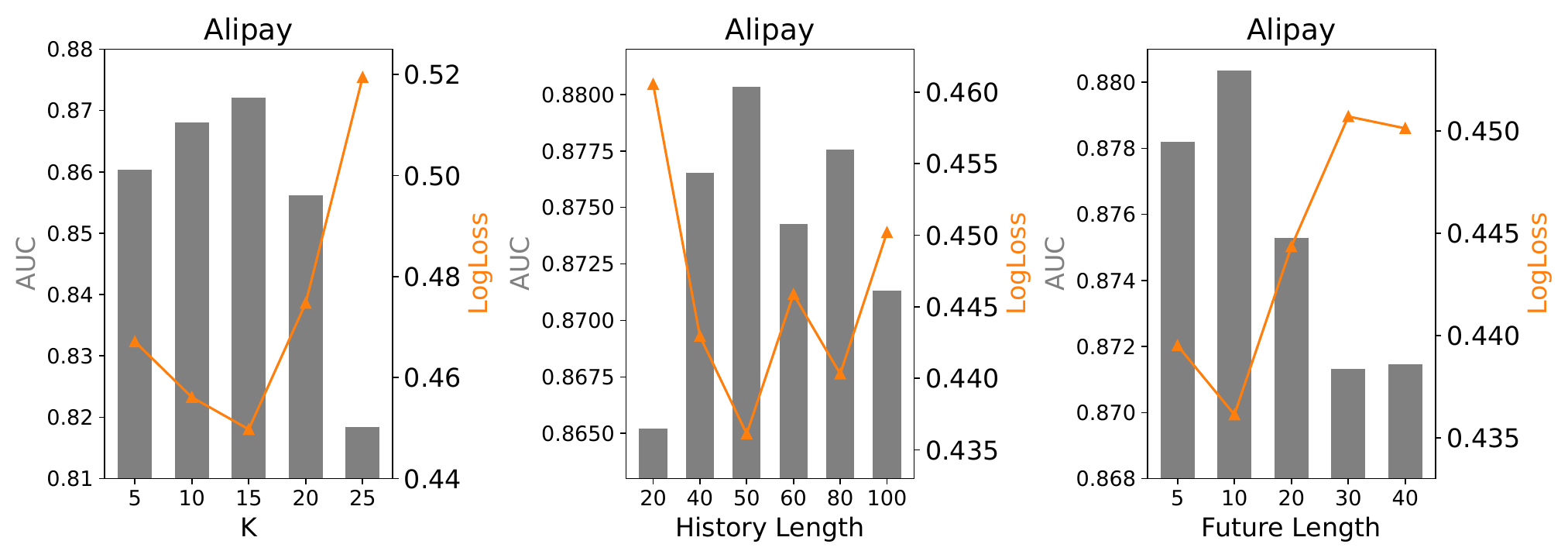}
% \vspace{-15pt}
\caption{Hyperparameters study of LIFT on Alipay.}%\vspace{-5pt}
\label{liftfig:alipay_hyper}
\vspace{-5pt}
\end{figure}

% \minisection{Hyperparameter in the Retriever  (RQ4)} 
In the LIFT framework, $L$ means the context sequence length and $K$ means the retrieved samples count. From Figure \ref{liftfig:taobao_hyper} and Figure \ref{liftfig:alipay_hyper}, we can find that with the increase of $K$, the final prediction result first goes up and then down. The curve indicates that along with with the $K$, the information first increases, and then more noises are introduced to become the dominant part. The AUC curves of $L$ show the similar trends of $K$, i.e., as $L$ increases, the AUC first gets better and then drops down.
We think it is the same reason as for $K$.
Moreover, because of the limitation of the GPU resources, we only conduct this hyperparameter study on Taobao and Alipay. and we could only increase the $L$ to 70 in the history length of Taobao, where the downtrend just started.

\end{document}